\documentclass[times]{aa}
\usepackage{psfig}
\usepackage{natbib}
\usepackage{graphicx}

\bibpunct{(}{)}{;}{a}{}{,}

\begin{document}

\def\eps{\varepsilon}
\def\aap{A\&A}
\def\apj{ApJ}
\def\apjl{ApJL}
\def\mnras{MNRAS}
\def\aj{AJ}
\def\nat{Nature}
\def\aaps{A\&A Supp.}
\def\prd{Phys. Rev. D}
\def\prl{Phys. Rev. Lett.}
\newcommand{\vt}{\mbox{\bf {T}}}
\newcommand{\vtcmb}{\mbox{\bf {T}}_{cmb}}
\newcommand{\vm}{\mbox{\bf {M}}}
\newcommand{\vn}{\mbox{\bf {N}}}
\newcommand{\vf}{\mbox{\bf {F}}}
\newcommand{\1}{\'\i}

\def\me{m_\e}
\def\lesssim{\mathrel{\hbox{\rlap{\hbox{\lower4pt\hbox{$\sim$}}}\hbox{$<$}}}}
\def\gtrsim{\mathrel{\hbox{\rlap{\hbox{\lower4pt\hbox{$\sim$}}}\hbox{$>$}}}}

\def\vr{\vec{r}}
\def\vrp{\vec{r}_\perp}

\def\del#1{{}}

\def\C#1{#1}

\input epsf
\def\plotancho#1{\includegraphics[width=17cm]{#1}}

\title{Spectral Indications of thermal Sunyaev-Zel'dovich Effect in {\sc Archeops} and
WMAP Data}
\titlerunning{Spectral Indications of tSZ in {\sc Archeops} and WMAP CMB Data}
\author{
C. Hern\'andez-Monteagudo \inst{1,2}  \and J.F. Mac\'{\i}as-P\'erez \inst{3} \and 
M. Tristram \inst{3} \and F.-X. D\'esert\inst{4}
}
\offprints{carloshm@astro.upenn.edu, macias@lpsc.in2p3.fr}
\institute{
Department of Physics \& Astronomy, University.of Pennsylvania,
209 South 33rd Str., Philadelphia, PA 19106, USA \and
Max-Planck-Institut f\"{u}r Astrophysik,
Karl-Schwarzschild-Str.1, Postfach 1317, 85741 Garching, 
Germany \and
Laboratoire de Physique Subatomique et Cosmologie, 53 av. des Martyrs, 38026 Grenoble CEDEX, France\and
Laboratoire d'Astrophysique, Obs. de Grenoble, BP 53, 38041 Grenoble Cedex 9, France
} 
\date{\today}

\abstract{
In this paper, we present a joint cross-correlation analysis of the {\sc Archeops} CMB maps at
143 and 217 GHz and the WMAP CMB maps at 41, 61 and 94 GHz
with sky templates of projected galaxy density constructed from the 2MASS Extended Source catalog. 
These templates
have been divided in patches sorted in decreasing galaxy density with a fixed number of pixels
(we considered patches having 32, 64 and 128 pixels) and the cross correlation has been performed
independently for each of these patches. We find that the densest
patch shows a strong temperature decrement in the Q, V, W bands of WMAP and in the 143 GHz channel
of {\sc Archeops}, but {\em not} in the 217 GHz channel. Furthermore, we find that the 
spectral behavior of the amplitude of this decrement is compatible with that expected for the
non-relativistic thermal Sunyaev-Zel'dovich effect, and is incompatible (at 4.5$\sigma$ level) with
the null hypothesis of having only CMB, noise and a dust component ($\propto \nu^2$) in those pixels. 
We find that the first 32-pixel sized patch samples the cores of 11 known massive nearby galaxy clusters. 
Under the assumption that the decrement
found in this patch is due entirely to the thermal Sunyaev-Zel'dovich effect, we obtain an average
comptonization parameter for those sources of $y = (0.41 \pm 0.08) \times 10^{-4}$ at 13 arcmin angular 
scales. This value is compatible at 1-$\sigma$ with the expectation,  $y = 0.49 \times 10^{-4}$,  
from a model of the cluster flux
number counts based on the standard $\Lambda$-CDM cosmology \cite{xue}.
The observed value of $y$ is slightly diluted when computed for the first patch of 64 and 128 pixels, 
presumably due to the inclusion of less massive clusters, and the dilution factor inferred is also compatible
with the quoted model.

 } \maketitle

\section{Introduction\label{sec:intro}}

The thermal Sunyaev-Zel'dovich effect (hereafter tSZ, \cite{tSZ})
constitutes a unique tool to explore the presence of baryons in the
Universe. It arises as a consequence of the distortion that the black
body spectrum of the Cosmic Microwave Background (CMB) radiation
experiences when it encounters a hot electron plasma. In this Compton
scattering, electrons transfer energy to the CMB radiation, generating
an {\em excess} of high energy photons and a deficit in the low energy
tail of the distribution. This photon reallocation translates into a
frequency dependent change of the brightness temperature of the CMB,
which, in the non-relativistic limit, has a very simple form, ($f_{tSZ}(x) = x /\tanh [x/2] - 4$,
with $x= h\nu / k_B T_{CMB}$ the adimensional frequency in terms of the CMB temperature 
monopole $T_{CMB}$). 
The amplitude of this distortion is proportional to the electron pressure
integrated along the line of sight ($\delta T_{tSZ} / T_{CMB} = f_{tSZ}(x) \int dr\; \sigma_T n_e \;
k_B T_e / (m_e c^2) $, with $\sigma_T$ the Thomson cross section and $n_e$, $T_e$ and $m_e$ the electron
number density, temperature and mass, respectively); and this makes this effect
particularly sensitive to collapsed or collapsing structures
containing hot electrons,
such as clusters and superclusters of galaxies (see \cite{Birk99} for a extensive review).\\

In addition to the intrinsic energy inhomogeneities of CMB radiation
generated during inflation, there are further temperature anisotropies
introduced in the CMB during recombination, which are mainly caused
by two physical processes. These processes are the {\it last} 
Doppler kick exerted by electrons via Thomson scattering just before 
recombining, and the subsequent gravitational redshift
experienced by CMB photons as they climb the potential wells generated
by the inhomogeneities in the matter distribution (Sachs \& Wolfe effect,
e.g. \cite{HuS95}). While all this happens at
$z\sim 1100$,  a similar scenario can take place at much lower
redshifts: as the first stars reionise the universe, new free
electrons are produced which again scatter CMB photons, partially
blurring primordial anisotropies generated during recombination and
introducing new ones at much larger angular scales. Also, if
$\Omega_{\Lambda}$ is non-zero, the decay of gravitational potentials
in linear scales introduces a net blueshift in the CMB radiation at
late epochs (z $<$ 2), which is known as the Integrated Sachs Wolfe
effect, (ISW \cite{ISW}). Despite the fact that most of these
phenomena introduce temperature fluctuations of amplitudes larger than
the tSZ effect, the particular frequency dependence of the latter
should enable its separation. Whereas the first generation of CMB
experiments like COBE \cite{cobe} and  Tenerife \cite{tenerife}
aimed to simply detect the largest CMB temperature anisotropies in the
big angular scales, experiments like, e.g., Boomerang \cite{boom}, VSA \cite{vsa},
{\sc Archeops} \cite{archeops} and WMAP \cite{wmap} have already reached the
sensitivity and angular resolution levels required to probe relatively
weak signals like the tSZ effect.

In this work we perform a combined analysis of {\sc Archeops} and WMAP
CMB data, searching for {\em spectral} signatures of the tSZ effect.
Previous works \cite{wmap, jalsc, myers, lettchm, afsh04,fosalba1,fosalba2} have
claimed the detection of tSZ in WMAP data at different significance
levels. However, all those studies were exclusively based on spatial
cross-correlations of large scale structure catalogues with CMB data.
In this work, we take advantage of the frequency coverage provided by
the combination of {\sc Archeops} and WMAP experiments in order to
include an analysis of the frequency behavior of a signal which is
spatially correlated with regions hosting large galaxy overdensities.\\

The sketch of the paper is as follows: in Section~\ref{sec:archeops}
we summarize the outcome of the {\sc Archeops} and WMAP experiments, whose data
products are analyzed as explained in Section~\ref{sec:analyss}.
Section~\ref{sec:res} shows our results, which are compared with those
obtained from WMAP data. Their implications are discussed in
Section~\ref{sec:discuss}. Finally, we conclude in Section~\ref{sec:concl}.\\

\section{The {\sc Archeops} and WMAP data set\label{sec:archeops}}

The {\sc Archeops} \cite{archeops}\footnote{see {\tt
    http://www.archeops.org}} experiment was designed to obtain a
large sky coverage of CMB temperature anisotropies in a single balloon
flight at millimeter and submillimeter wavelengths. {\sc Archeops} is
a precursor to the {\sc Planck HFI} instrument \cite{lamarre}, using
the same optical design and the same technology for the detectors,
spider--web bolometers, and their cooling 0.1~K dilution fridge. The
instrument consists of a 1.5~m aperture diameter telescope and an
array of 21~photometric pixels operating at 4~frequency bands centered
at 143, 217, 353 and 545~GHz. The two low frequencies are dedicated to
CMB studies while high frequency bands are sensitive to foregrounds,
essentially to interstellar dust and atmospheric emission.
Observations are carried out by spinning the payload around its
vertical axis at 2~rpm. Thus the telescope produces circular scans at
a fixed elevation of $\sim 41$~deg. The data were taken during the
Arctic night of February~7,~2002 after the instrument was launched by
CNES from the Esrange base near Kiruna (Sweden). The entire data set
covers $\sim 30$\% of the sky in 12 hours of night observations.

For the purpose of this paper, we concentrate in the low frequency
channels at 143 and 217 GHz. Maps for each of the bolometers have been
produced from the {\sc Archeops} processed and foreground cleaned
timelines, using the Mirage optimal map making code \cite{mirage} as
discussed in \cite{archeops_cl2}.
The maps for the 4 most sensitive bolometers at 143 GHz were combined
into a single map at 143 GHz and equally the two most sensitive
bolometer maps were combined at 217 GHz. The CMB dipole is the prime
calibrator of the instrument. The absolute calibration error against
the dipole as measured by COBE/DMR \cite{fixsen} and confirmed by
{\sc WMAP} \cite{wmap} is estimated to be 4\% and 8\% in temperature
at 143~GHz and 217~GHz respectively.  These errors are dominated by
systematic effects.  The noise contribution in the combined maps at
143 and 217 GHz was computed using Monte Carlo simulations. For each
bolometer, by using the power spectrum of the noise in the time domain
data set, we produced fake timelines of {\sc Archeops} noise. These
were processed and projected into maps following the same procedures
used for the {\sc Archeops} data themselves as described before. \\

The WMAP satellite mission was designed to measure the CMB 
temperature and polarization anisotropies in
5 frequency bands, 23, 33, 41, 61 and 94 GHz with a full sky coverage.
The satellite was launched in June 2001 and its first results 
after the first year of observations \cite{wmap} included the CMB temperature
and temperature-polarization cross-correlation power spectra, as well as
full sky temperature maps for each of the frequency bands. 
In this paper we consider only data from the high frequency channels,
Q [41 GHz], V[61 GHz] and W [94 GHz], since only for these bands there
are foreground clean maps available at the LAMBDA site {\tt http://lambda.gsfc.nasa.gov/}.



\section{The Statistical Analysis\label{sec:analyss}}

In this Section we outline the correlation analysis performed on
{\sc Archeops} and WMAP data and the 2MASS Extended Source Catalog (XSC, \cite{Jarrett})
on the sky region covered by {\sc Archeops}. 
This analysis
is essentially identical to that applied in \cite{lettchm}, 
paper to which we refer for a more detailed description of the
statistical method. It consists of a pixel-to-pixel comparison of
{\sc Archeops} CMB data with a template of the large scale structure built from 
the 2MASS XSC catalog. \\

The 2MASS XSC catalog contains approximately 1.6
million galaxies detected in the infrared filters I, J and K, and covers
the whole celestial sphere. Those frequencies are particularly insensitive to 
dust absorption, and for this reason this catalog can trace the extragalactic
structure at very low galactic latitudes.
The galaxy templates built from it take into account the spatial
distribution of the galaxies and the instrumental beam of the CMB
experiment. By using the HEALPix\footnote{HEALPix's URL site: \\
{\tt http://www.eso.org/science/healpix/}} \cite{healpix} tessellation of the sky, 
we built a map with the same resolution parameter ($N_{side} = 512$) than the 
one used in {\sc Archeops} and WMAP data. Every pixel was assigned a value equal to the
number of galaxies present in such pixel. The resulting map was then convolved
with a window function corresponding to the instrumental beam of each 
of the detectors taken into consideration. 
For {\sc Archeops}, the resulting templates were then
weighted by their noise levels and co-added per frequency band, in such a way that we ended up
with two different galaxy templates corresponding to the 143 GHz and 217 GHz 
bands. For WMAP, we produced templates for the Q, V and W bands.\\

In the next step, we sorted the pixels of each template in terms of
its amplitude, so that {\em first} pixels would have {\em higher}
galaxy densities. These pixels were grouped in patches of varying
sizes (32, 64 or 128 pixels per patch), and again patches were sorted
in such a way that first patches contained larger projected galaxy
densities.  Next, we analyzed {\em each} of these patches separately,
by comparing them to the corresponding patches in the
{\sc Archeops} and WMAP CMB
maps on a pixel-to-pixel basis. 

As explained in e.g. \cite{jalsc}, it
is possible to estimate the contribution of a given spatial template
(\vm) on a total measured temperature map (\vt), which is the result
of the addition of several components:
\begin{equation}
\vt = \vtcmb + {\tilde \alpha}\cdot\vm + \vn,
\label{eq:tmodel}
\end{equation}
namely CMB ($\vtcmb$), instrumental noise ($\vn$)
and some signal coming from the extragalactic template ($\vm$). 
The contribution of $\vm$ is parametrized by ${\tilde \alpha}$, 
and an optimal value of it
({\it optimal} in terms of the temperature model given in eq.(\ref{eq:tmodel})),
together with its formal error bar, is given by
\begin{equation}
\alpha = \frac{ \vt {\cal C}^{-1} \vm^T} {\vm {\cal C}^{-1} \vm^T },
\;\;\;\; \sigma_\alpha = \sqrt{\frac{1}{\vm{\cal C}^{-1}\vm^T}}.
\label{eq:alpha1}
\end{equation}
The matrix ${\cal C}$ is the covariance matrix of \vt, which must
contain the correlation matrices of both the CMB and noise components.
For the small scales we are probing here (the pixel has a size of
$\sim 7$ arcmins), the noise is the main contributor to the covariance
matrix\footnote{Note that for these small scales, the assumption of a
  piossonian noise is a good approximation}. In our case, this matrix
must be evaluated {\em only for the pixels belonging to the patch
  under analysis}. Since our patches are relatively small, the
inversion of this matrix poses no numerical problem.\\

Therefore, for every patch a value of $\alpha$ and $\sigma_{\alpha}$
was obtained. However, uncertainties in the determination of the noise
amplitude may bias our determination of $\sigma_{\alpha}$, and for
this reason we computed a different estimate of the uncertainty of
$\alpha$, namely the r.m.s variation of this parameter for all available
patches, which will be denoted as
$\sigma_{\alpha}^{rms}$. We remark the fact that,
according to eq.\ref{eq:alpha1}, an error in the noise normalization
does not affect the estimates of $\alpha$, but only those of
$\sigma_{\alpha}$.  
When comparing $\sigma_{\alpha}$ with $\sigma_{\alpha}^{rms}$, we found that
for the 143 GHz channel of {\sc Archeops}, the latter was about a 40\% larger than the former, 
and hence we decided to adopt it in order to quote conservative estimates of
statistical significance. For the 217 GHz case no such bias was found, and we decided to use
again $\sigma_{\alpha}^{rms}$.


\section{Combined results for {\sc Archeops} and WMAP\label{sec:res}}

\begin{figure*}
\centering
\plotancho{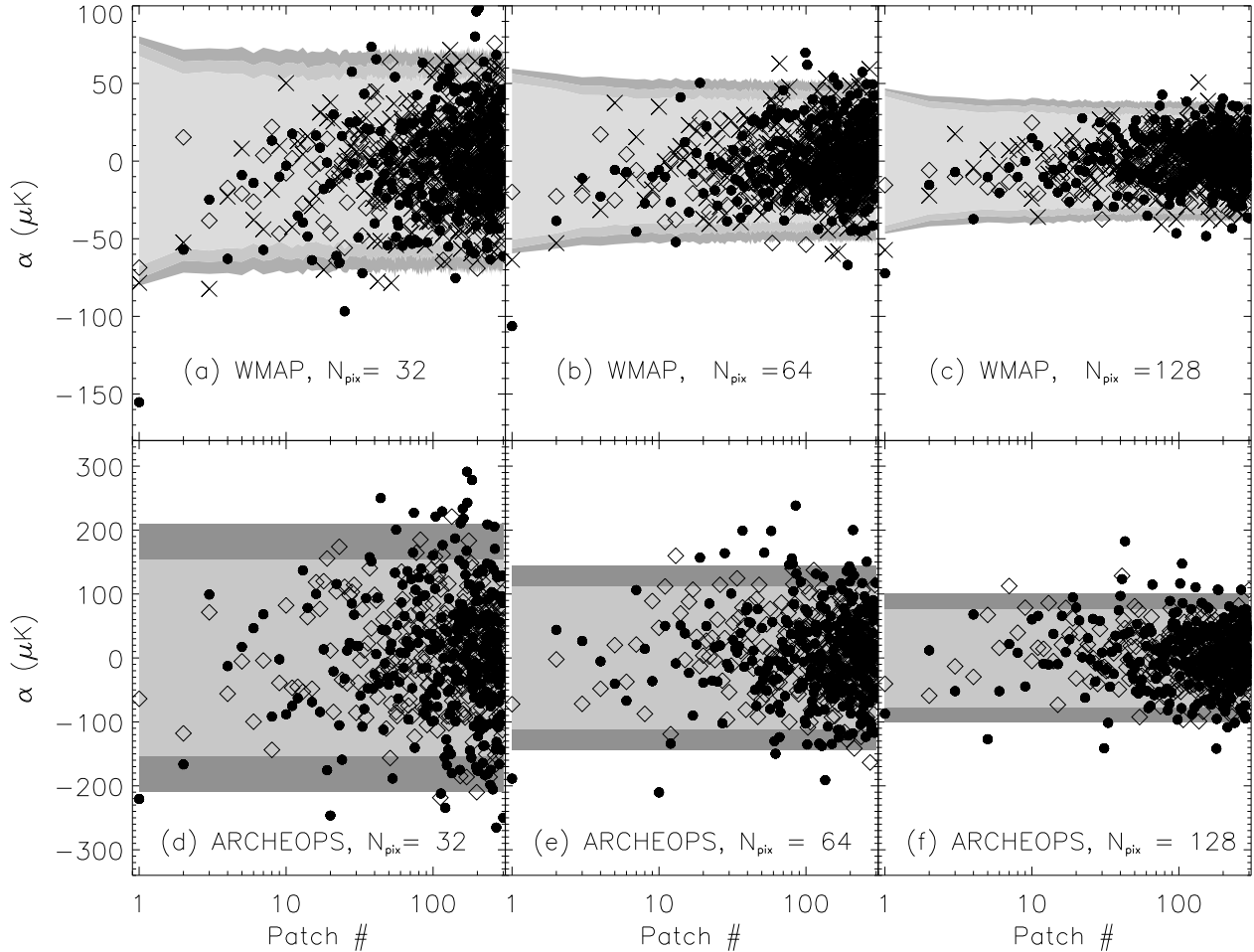}
\caption[fig:figure1]{$\alpha$'s obtained by means of eq.(\ref{eq:alpha1})
 from {\sc Archeops} and WMAP data in the fraction of the sky covered by the former
experiment. In panels (a-c), filled circles, crosses and diamonds refer
to the W, V and Q bands respectively. In panels (d-f), the 143GHz and 
217GHz channel results are given by filled circles and diamonds, respectively.
The colored bands display the 2-$\sigma$ 
confidence limit for the $\alpha$'s: dark grey for filled circles, intermediate
grey for crosses, and light grey for diamonds. 
Our analysis are focused on the first patch, hosting the highest
projected galaxy density.}
\label{fig:figure1} 
\end{figure*}

In figure \ref{fig:figure1} we plot the recovered $\alpha$'s versus
the patch index. Results are grouped in six different panels: top and 
bottom panels refer to WMAP and {\sc Archeops} experiments, respectively, 
whereas left, middle and right panels display results for patches of 32, 
64 and 128 pixels, respectively. For {\sc Archeops}, filled
circles and diamonds refer to 143 GHz and 217 GHz respectively,
whereas for WMAP those symbols correspond to the W and the Q channels, being
the results of the V band given by the crosses.
Dark and light grey colored bands limit the 2-$\sigma$ confidence
levels for filled circles and diamonds, respectively, while the moderately
dark bands refer to the V band (crosses) in the case of WMAP. 
As explained above, for the 143 GHz and 217 GHz channels
the amplitude of the shaded regions was computed from
the typical dispersion of the values obtained for $\alpha$ in patches 
where the tSZ contribution is expected to be negligible, i.e., 
in patches with indexes between 40 and 300.
We have found that for the 143 GHz channel the first patch contains an unusual
negative $\alpha$ while at 217 GHz seems to be compatible with zero.  This patch, in the case it contains 64 pixels,
hosts the central pixels of 20 different ACO clusters of galaxies,
COMA among them. Out of them, 11 (COMA again included) are already
sampled by the 32 densest pixels.
Its statistical significance is slightly bigger for patches with 64
pixels ($>$ 2.5-$\sigma^{rms}$), since it contains the first two {\it
  very negative} patches of 32 pixels each, (see figure
\ref{fig:figure1}d).  In figure \ref{fig:figure1}e, from the first 300
patches, very few of them ($\sim 12$) depart from zero by an amount
similar to that of the first patch; and such number is very close to
what one expects under Gaussian statistics at 2.5-$\sigma$ level of
significance. This peculiar behavior of the first (or first two)
patch(es) disappears at 217 GHz (see diamonds): in no case the
diamonds corresponding to the first two patches trespass the
2-$\sigma^{rms}$ limit. This picture is consistent with part of the
signal being generated by the tSZ effect, since such component is
negative at 143 GHz and becomes zero at 217 GHz.\\

In order to interpret the results from WMAP, one musts keep in mind that
the Q and V bands have remarkably larger beams than the W band: 
while the {\sc Archeops}'
and W band's Point Spread Functions are similar in size 
($\approx 13$ arcmins), the
beams of the Q and V bands have an average (linear) size of $\sim$ 31 and 21
 arcmin respectively.
This, in terms of tSZ flux, corresponds to factors $\sim$ 5.7 and 2.6 smaller 
in the Q and V band for point-like objects. On the other hand, it is clear
from figure \ref{fig:figure1} that WMAP has a much lower noise level when
compared to {\sc Archeops}, (approx. a factor of 3.5). As one looks at the
top panels in figure \ref{fig:figure1}, one finds that 
for the first patch of both 32 and 64 pixel size, the W band gives a 
decrement about 4-$\sigma$ away from zero, (which however is not found
for the second patch of 32 pixel size). This statistical singularity 
of the first patch decreases remarkably in the Q and V bands, 
but still remains close to the 2-$\sigma$ level. However, due to the 
argument on beam dilution on bands Q and V with respect to W, we still find
the amplitudes given by the three bands of WMAP consistent with being 
(at least partially)
generated by unresolved objects causing a temperature decrement. 

It also worth to remark that, if the first patch contains pixels
which do {\em not} contribute considerably to the tSZ decrement, the
significance of the overall $\alpha$ obtained for that patch will
diminish accordingly. This is our explanation for the decreasing
significance of the values of $\alpha$ in the first patch for a size
of 128 pixels when compared to a size of 64 pixels.


\section{Discussion\label{sec:discuss}}

\subsection{Spectral dependence of the cross-correlation coefficients}

\begin{figure}[t]
\begin{center}
\psfig{figure=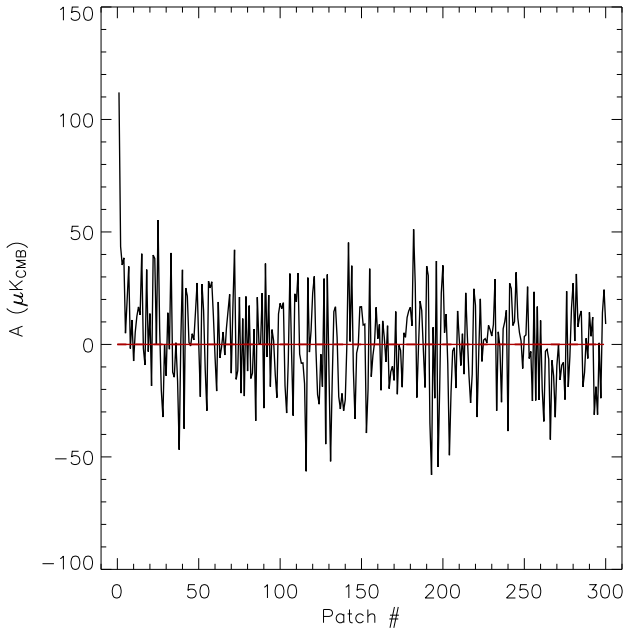,width=0.5 \textwidth,angle=0}
\end{center}
\vspace{-0.8cm}
\caption[]{tSZ fitted amplitude, $A$, as a function of patch number. We only observe a significant temperature decrement
($A > 0$) in patch number 1. \label{fig:A}}
\end{figure}

As discussed in the previous section, Fig.~\ref{fig:figure1} indicates
a significant temperature decrement 
in patch 1 for
both the {\sc Archeops} and WMAP data sets which seems to be compatible with 
the tSZ effect. For a better assessment of this result we have
compared, via a linear fit, the observed spectral dependency of the
correlation coefficients $\alpha$ in each of the patches to the
following model
\begin{equation}
\alpha(\nu) = A \times f_{tSZ}(\nu) + n(\nu)
\end{equation}
where $A$ is global calibration factor which is estimated from the fit, $f_{tSZ}(\nu)$ represents the 
spectral behavior of the tSZ-induced change in brightness temperature 
and $n(\nu)$ is the instrumental noise in $\alpha(\nu)$.
For this model if the signal observed is compatible with the tSZ effect 
we expect a temperature decrement $A$ to be significantly positive, otherwise we expect 
no correlation and therefore $A$ to be compatible with zero. 
We must remark that, due to the different size of the beams for every channel,
we had to rescale the $\alpha$'s to a common reference beam size of 13 arcmin. When doing this,
we assumed that the signal was generated by unresolved sources, and hence we scaled the
$\alpha$'s as the ratio of the area of the beam with the reference one.

\begin{figure}[t]
\begin{center}
\psfig{figure=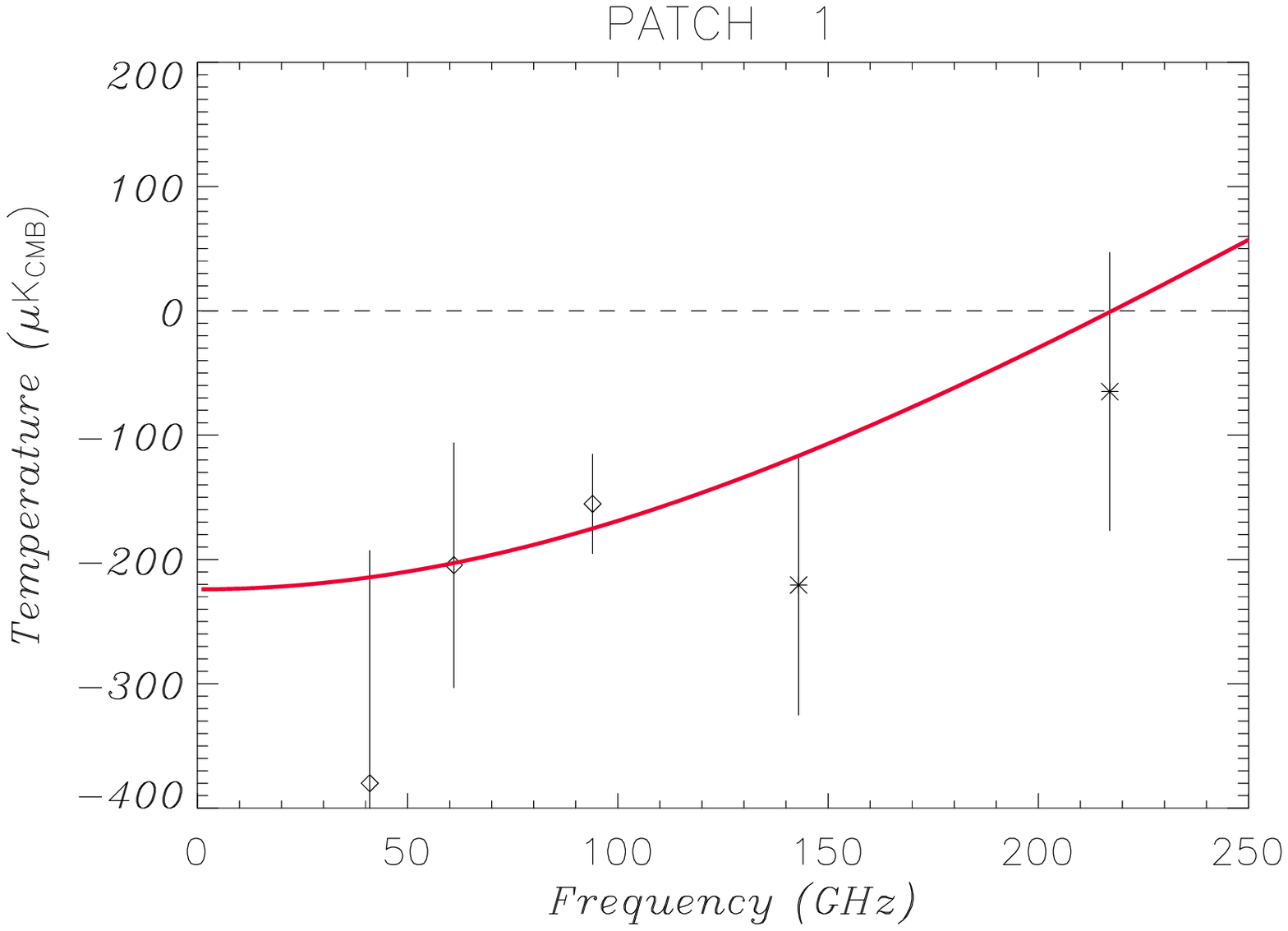,width=0.5 \textwidth,angle=0}
\psfig{figure=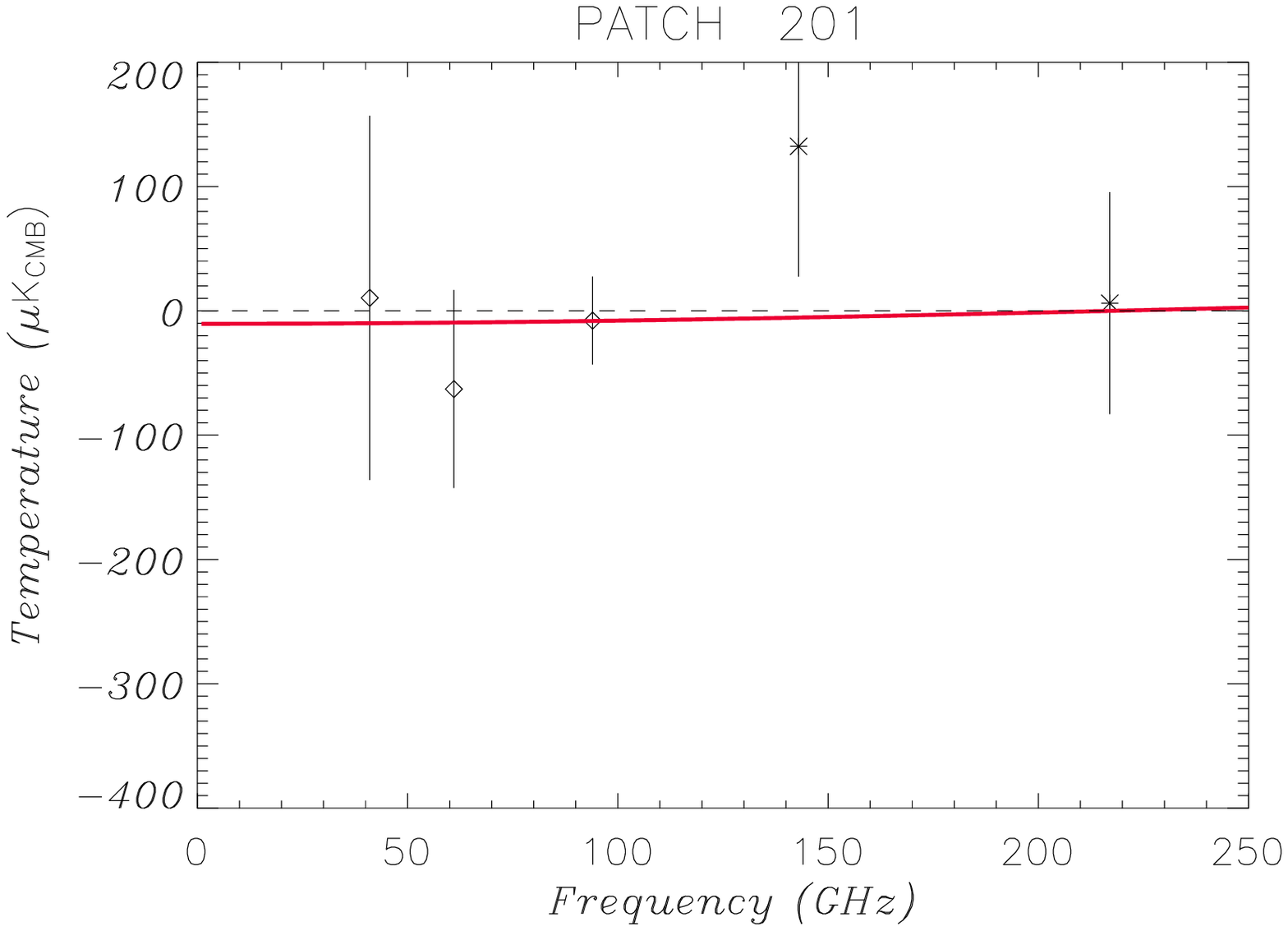,width=0.5 \textwidth,angle=0}
\end{center}
\vspace{-0.8cm}
\caption[]{The $\alpha$ correlation coefficient in CMB temperature units as function of frequency in GHz for patch
  1 (top panel) and patch 201 (bottom panel) of the 32 pixel patches.
  The diamonds are the WMAP data points and the stars the {\sc
    Archeops} one. In red, we overplot the best-fit $tSZ$ computed as
  discussed in the text.  \label{fig:fig2} }
\end{figure}

\begin{figure}[t]
\begin{center}
\psfig{figure=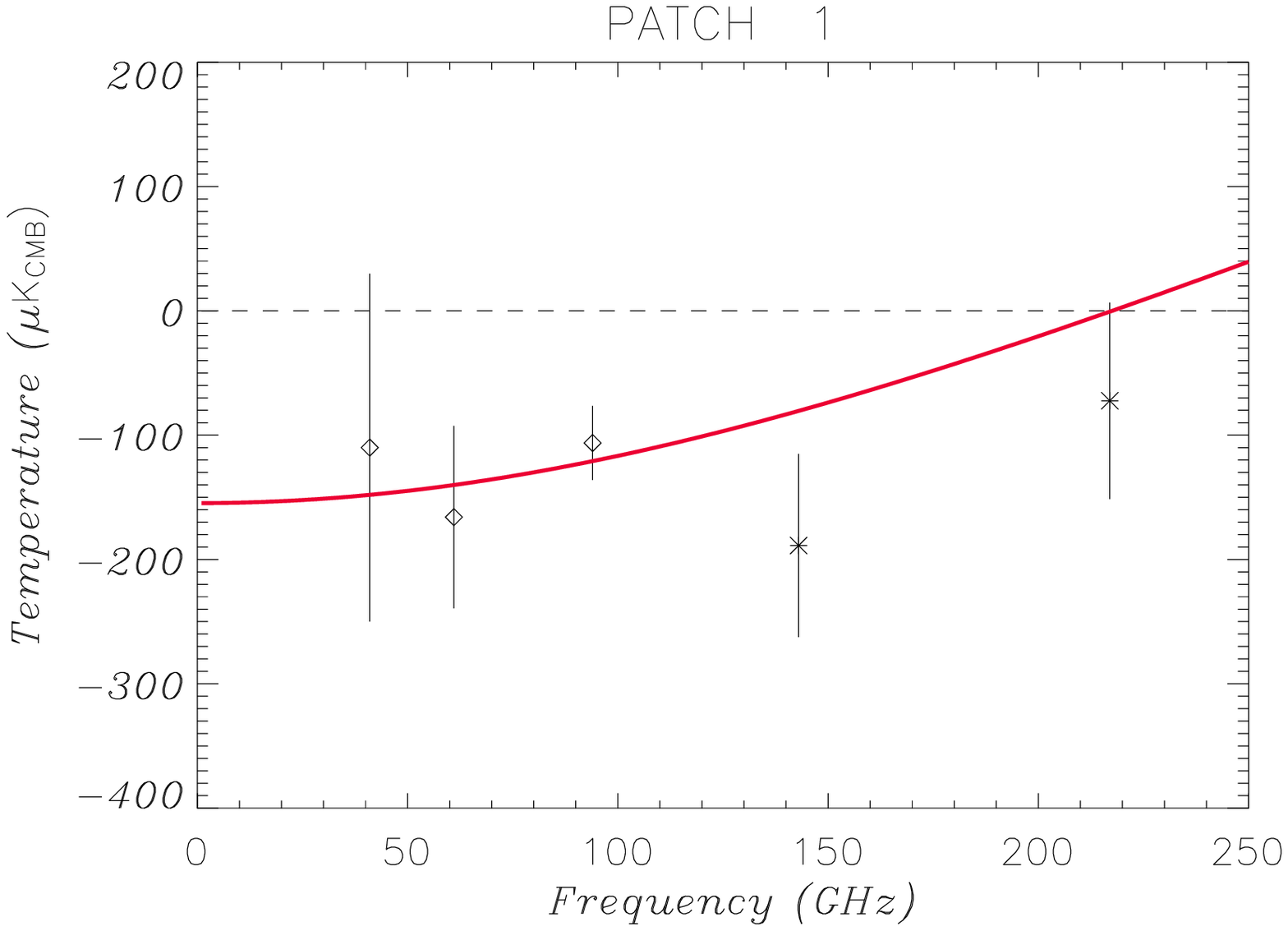,width=0.5 \textwidth,angle=0}
\psfig{figure=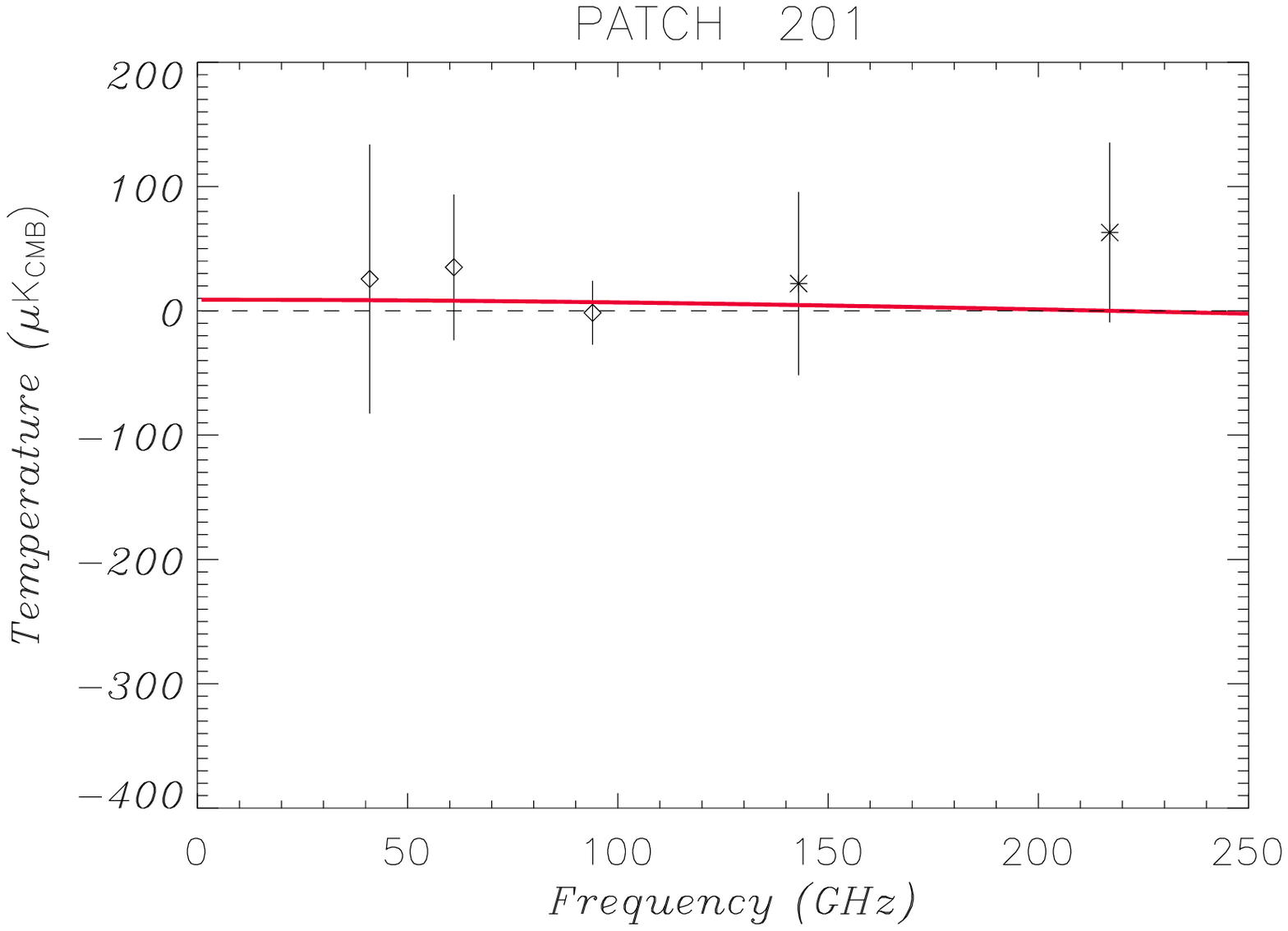,width=0.5 \textwidth,angle=0}
\end{center}
\vspace{-0.8cm}
\caption[]{$\alpha$ correlation coefficient in CMB temperature units as function of frequency in GHz for patch
1 (top panel) and patch 201 (bottom panel) of the 64 pixel patches. The diamonds are the
WMAP data points and the stars the {\sc Archeops} one. In red, we overplot the best-fit $tSZ$ computed as
discussed in the text. \label{fig:fig3} }
\end{figure}

\begin{figure}[t]
\begin{center}
\psfig{figure=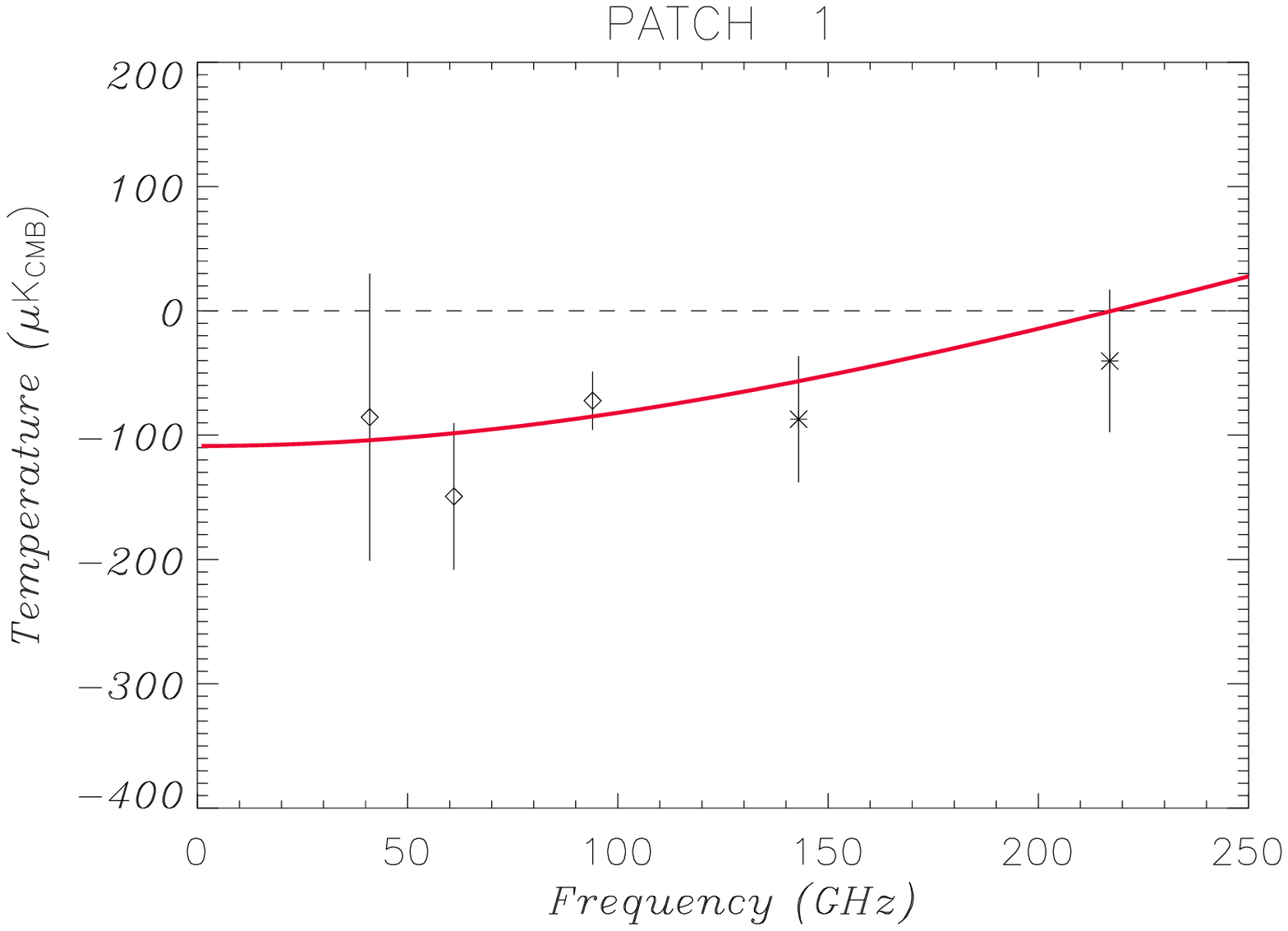,width=0.5 \textwidth,angle=0}
\psfig{figure=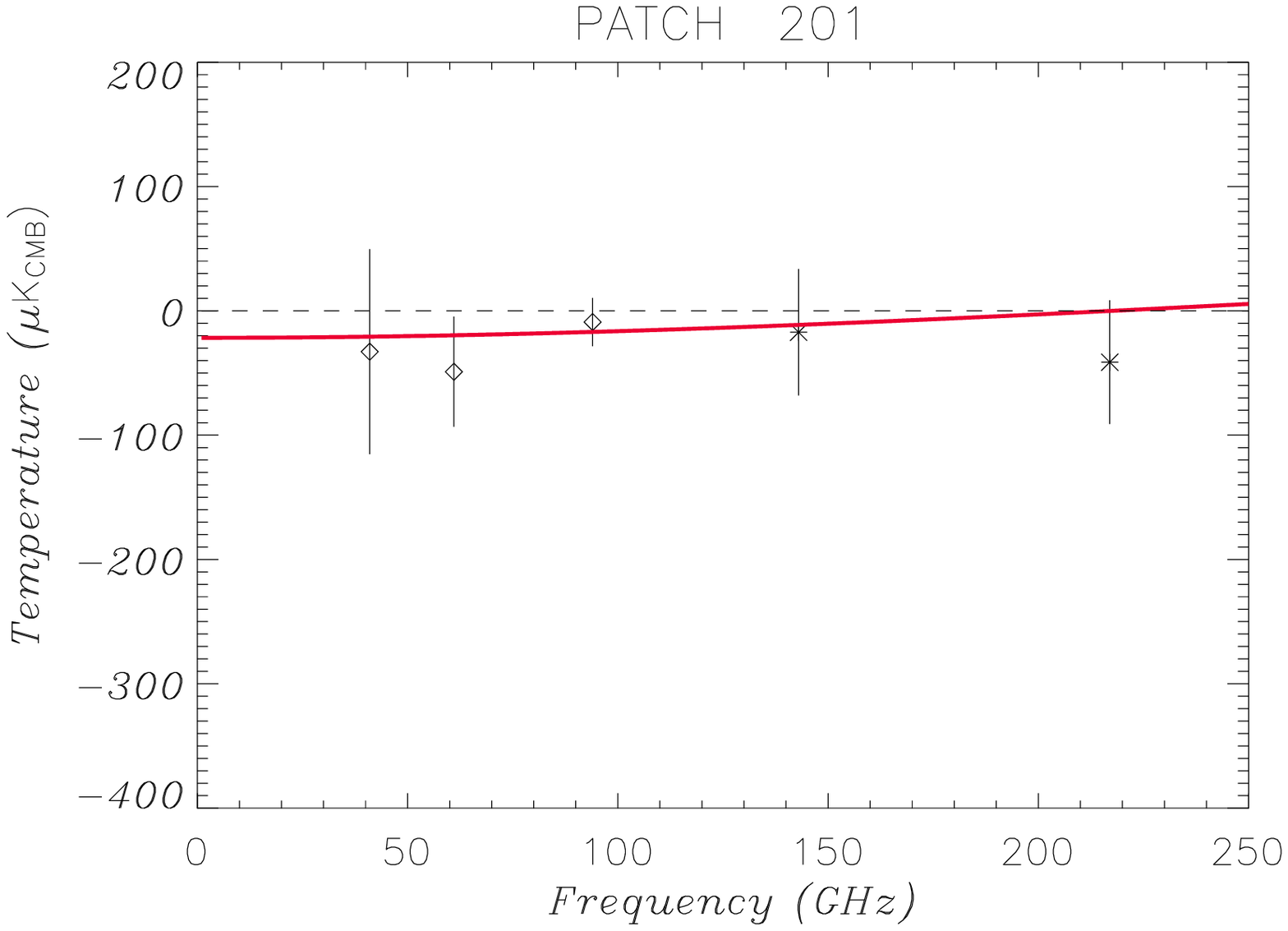,width=0.5 \textwidth,angle=0}
\end{center}
\vspace{-0.8cm}
\caption[]{The $\alpha$ correlation coefficient in CMB temperature units as function of frequency in GHz for patch 
1 (top panel) and patch 201 (bottom panel) of the 128 pixel patches. The diamonds are the 
WMAP data points and the stars the {\sc Archeops} one. In red, we over plot the best-fit $tSZ$ computed as
discussed in the text. \label{fig:fig4} }
\end{figure}

The main results of the fitting procedure described before are shown on figure \ref{fig:A} where we trace
the fitted amplitude, $A$, of the tSZ signal as a function of the patch number for patches of 32 pixels. 
We observe that only for patch number 1, where most massive clusters might be, there is a significant 
temperature decrement ( $A>0$ ). 
For a closer look to the fits, figures \ref{fig:fig2}, \ref{fig:fig3}, \ref{fig:fig4} represent the
$\alpha$ coefficient in CMB temperature units as function of frequency in
GHz for the 32, 64 and 128 pixel patches respectively. The top plot
corresponds to patch 1 and the bottom one to a patch containing 
an average value of the projected galaxy density (patch 201).  
In the three figures we observe that the correlation
coefficients found for the patch 1 are consistent which what we expect
for tSZ emission with $A$ values of $112 \pm 22$ $\mu$K, $77 \pm 16$ $\mu$K and
$54 \pm 13$ $\mu$K respectively. Notice that there is good agreement between
the model and the data with reduced $\chi^2$ values of $3/4$, $
3.4/4$ and $2/4$ for patches containing 
32, 64 and 128 pixels respectively. For
the null hypothesis ($A=0$ $\mu$K), the reduced $\chi^2$ values are $28/5$,
$26/5$ and $20/5$. We interpret these results as a spectral indication
of the measurement of a tSZ signal in patch 1.  However, for all
other patches we find $A$ compatible with zero, showing that the data are in 
good agreement with the null hypothesis. For example, for patch 201 the $A$
values are $5 \pm 19$ $\mu$K, $-4 \pm 13$ $\mu$K and $10 \pm 10$ $\mu$K with reduced
$\chi^2$ values of $2.2/4$, $1.2/4$ and $1.3/4$ respectively.  
We consider these results compatible with a no
detection of tSZ signal in patch 201.

The uncertainties for the total amplitude of the tSZ signal, $A$, presented above do not account for systematic
errors on the {\sc Archeops} data. These are dominated by residuals from atmospheric and Galactic 
dust emissions which in a first order approximation increase as $\nu^2$ in the {\sc Archeops}
frequency range. To account for those contributions we have recomputed the total amplitude
for the tSZ signal adding an extra term in the fitted function as follows:
\begin{equation}
\alpha(\nu) = A \times f_{tSZ}(\nu) + A_{sys} \times \nu^2 +n(\nu),
\end{equation}
where $A_{sys}$ is computed only from the {\sc Archeops} data.
For patch 1 we obtain $A$ values of $110 \pm 22$ $\mu$K, $74 \pm 16$ $\mu$K and $53 \pm 12$ $\mu$K, 
and $A_{sys}$ values of $-95 \pm 102$ $\mu$K, $-104 \pm 71$ $\mu$K and $-46 \pm 51$ $\mu$K for patches 
of 32, 64 and 128 pixels respectively. From these results we conclude that the
$A$ coefficients are not significantly affected at 1-$\sigma$ level by systematic effects present 
in the {\sc Archeops} data.

\begin{table*}[h]
  \begin{center}
    \begin{tabular}{|c c c c c c |}\hline
      Npix & CMB Data & M1 & M2 & M3 & B\\
      \hline\hline
      32 &  $(0.4 \pm 0.08) $  &  $0.49 $  &    $3.91$   &   $6.24 $   &   $2.02 $\\
      64 &  $(0.27 \pm 0.06) $  & $0.31 $  &    $2.46 $   &   $3.50 $   &   $1.27 $ \\
     128 &  $(0.19 \pm 0.04) $  & $0.19 $  &    $1.55 $   &   $1.96 $   &   $0.80 $ \\
      \hline
    \end{tabular}
  \end{center}
  \caption{Expected average Compton parameter, $\langle y \rangle / 10^{-4}$, 
in patch 1, as a function of the number of pixels in the patch
for the four number counts models discussed in the text, M1, M2 and M3 \cite{xue} and B \cite{benson}}
  \label{table:Y}
\end{table*}

\subsection{Determination of the mean and integrated Compton parameters}

It is interesting to check whether pixels in patch 1 correspond to potential sources of tSZ or not.
For the 32-pixel-size patches, patch 1 includes
COMA, A0576, A0671, A0952, A1795, A2061, A2065, A2244, A2245, A2249, 
A2255. Since these are massive and relatively nearby galaxy clusters, it
is reasonable to expect some signature of the tSZ effect.
Using the quoted values for the $A$ parameter we infer the following estimate 
for the average Compton parameter in all those sources: $y = (0.41 \pm 0.08) \times 10^{-4}$.
In the case of 64-pixel-size patches we have roughly the same clusters and we expect
therefore the signal to be diluted, $y = (0.28 \pm 0.06) \times 10^{-4}$.
Finally for 128 pixel-size patches, patch 1 includes the following 27 clusters~: 
COMA, A0077, A0104, A0272, A0376, A0407, A0576, 
A0671, A0952, A1035, A1185, A1235, A1377, A1767, A1795, A1800, A2034, 
A2061, A2065, A2069, A2142, A2151, A2199, A2244, A2245, A2249, A2255. We deduce
for them $y = (0.20 \pm 0.05) \times 10^{-4}$.
Since we have scaled the $\alpha$'s to the beam-size of the 143 GHz channel of {\sc Archeops}, 
these quoted values of $y$ are associated to a (linear) angular scale on the sky of $\sim$ 13 arcmins.
For the above results we have assumed the CMB temperature to be $T_{CMB} = 2.725$K \cite{mather}. \\

We now try to relate the observed average tSZ decrement $\langle y_{obs} \rangle$ to the high end of
the SZ number counts. We make 2 hypotheses: 1) that the Archeops beam encloses
most of the integrated tSZ effect in clusters and 2) that the 2MASS survey is a
perfect tracer of the tSZ effect. We will come back to these hypotheses at the
end.

The correlation analysis, that is presented in the previous section, can be
recast in stating that the $N_{pix}$ brightest clusters of galaxies have an average
integrated $Y=\int d\Omega y$ parameter equal to $Y= \langle y_{obs}\rangle \Omega_{beam}$
where $\Omega_{beam}=2\pi (\rm{FWHM}/\sqrt{8\log{2}})^2$ is the Archeops beam solid
angle if $\rm{FWHM} = 13 \;\rm{arcmin}$. 
The integrated Compton parameter can be directly related to the SZ flux number counts
which has been the issue of a great number of studies (e.g. \cite{nabila, bartlett, barbosa,bartelmann}) from which we select \cite{xue} and \cite{benson}.

Following \cite{xue} there are
three possible models for which the number counts of clusters over the whole sphere can be parametrized
in terms of $Y$ as follows,

\begin{equation}
N(>Y)= N_0 (Y/Y_0)^{-\gamma},
\label{eq:nvsy}
\end{equation}

where we fiducially consider $Y_0= 10^{-2} \rm{arcmin}^2$

For Model 1 (M1), deduced from the cosmological $\Lambda$CDM matter power spectrum,
$N_0=29$ and $\gamma=1.5$. For Model 2 (M2), a non-evolving X-ray luminosity
function is used to correct the counts and gives a larger $N_0=635$ with the
same exponent. Finally Model 3 (M3), an evolving X-ray luminosity function is used
instead, giving $N_0=350$ with a flatter exponent $\gamma=1.2$. The conversion
from flux in Jansky units to the Compton frequency-independent quantity $Y$ is
obtained via

\begin{equation}
 \frac{F_{\nu}}{Y}=T_{CMB}\; \frac{\partial B_{\nu}}{\partial T_{CMB}}\; f_{tSZ}(\nu).
\label{eq:FoverY}
\end{equation}
For example, the fiducial value $Y_0= 10^{-2} arcmin^2$ is equivalent to 0.75
and 0.91~Jy at 90 and 143~GHz (resp.).

From the above formulas, we deduce the lower limit $Y_{min}$ in the number counts
for the first $N_{cl}= 32, 64, 128$ brightest clusters as

\begin{equation}
 Y_{min}= Y_0 \bigl( \frac{N_{cl}}{N_0 f_{sky}}\bigr)^{-1/\gamma}
\label{eq:miny} 
\end{equation}
where $f_{sky}\sim 0.20$ is the effective sky fraction used in the Archeops--WMAP
tSZ cross analysis. The average $Y$ value of these clusters is then

\begin{equation}
\langle Y \rangle= Y_0 \frac{\gamma}{\gamma-1} \bigl( \frac{N_{cl}}{N_0 f_{sky}}\bigr)^{-1/\gamma}
\label{eq:meany}
\end{equation}


We deduce the average Compton parameter for those clusters $\langle y \rangle$ as
\begin{equation}
\langle y \rangle =  \frac{\langle Y \rangle}{2\pi (\rm{FWHM}/\sqrt{8\log{2}})^2}
\label{eq:avy}
\end{equation}

For the previous three models, expected values for $\langle y \rangle$ are between $0.5\times
10^{-4}$ and $6\times 10^{-4}$. The value expected from basic
principles (M1), $\langle y \rangle = 0.49\times 10^{-4}$, is quite close to the
observed value $\langle y \rangle_{obs} =0.41\pm0.08\times 10^{-4}$. The observed
value should however be corrected up-wards. Non--linearities in the
relation between the 2--MASS density field and the $Y$ parameter
introduce an efficiency which is difficult to estimate, although we
note that as discussed above most of the brightest pixels in the 
density map constructed from the 2--MASS survey are
associated to well-known clusters of galaxies. 
If we considered that only
identified clusters (11) are present in the first patch of 32 pixels, the
above expected values should be multiplied by a factor of two.
Extension of the bright
clusters beyond the fiducial beam of 13~arcmin (like Coma) may produce a
differential effect with the observing beam at different frequencies,
as well as a bias in the number counts.

We also note that the dependence of $\langle y \rangle$ with the number of pixels
taken in the analysis (32, 64 or 128) follows quite well the Model 1
prediction as shown in Table~{\ref{table:Y}. We have also cross-checked the 
\cite{xue} modeling with that from \cite{benson}. This alternative model produces an intermediate
value between models M1 and M2 and is marginally compatible with the
observations.

\section{Conclusion\label{sec:concl}}

In this paper we have presented a joint cross-correlation analysis
of the {\sc Archeops} and WMAP data sets 
with a template of galaxy density constructed from the 2MASS XSC galaxy catalog. We have first divided the 
2MASS sky density map in patches of equal number of pixels and sorted these
patches in terms of decreasing 
projected galaxy density. For each of these patches
we have performed a cross-correlation analysis with the {\sc Archeops} data at 143 and 217 GHz and with the
WMAP data for the Q, V and W bands. For patches containing the densest 32, 64 and 128 pixels (patch 1), the
correlation test pointed to a prominent temperature decrement in WMAP's Q, V and W bands 
 and in the 143 GHz band of {\sc Archeops}, but {\em not} at 217 GHz, 
as would be expected for tSZ-induced temperature fluctuations. All the other patches failed to show a 
similar behavior. \\


To corroborate these results, for each of the patches we have compared the 
cross-correlation coefficients to a model of the tSZ frequency pattern, in which we 
fix the spectral behavior and fit for a global amplitude parameter $A$.
For the first patch and the three sizes considered (32, 64 and 128 pixels), 
we obtain non zero $A$ values at more than 4.5-$\sigma$ level with good 
agreement between the model and the data, and negligible contribution
from systematic residuals in the Archeops data. For all other pixels having smaller projected
galaxy density, we fail to detect any signature of tSZ effect.

From these results, we conclude that there is clear indication
of tSZ effect for patch 1 in the {\sc Archeops} and WMAP data sets. 
This is not surprising, since patch 1 contains pixels centered in 
massive and relatively nearby galaxy clusters. Assuming the signal observed is tSZ,
we infer, for the 32-pixels case, an average value for the comptonization parameter of 
$y= (0.41 \pm 0.08) \times 10^{-4}$ in all those clusters at an angular scale of $\sim$ 13 arcmins.
This value is compatible at 1-$\sigma$ level with the expectations,  $y = 0.49 \times 10^{-4}$ (cf. Table~\ref{table:Y}),  
from a model of the cluster flux number counts based on the standard $\Lambda$-CDM model, M1, assuming 
the measured $y$ is due to the contribution from the 32 brightest clusters. 
For 64 and 128-pixeles patches the tSZ signal is diluted, probably due to the contribution of relatively not so massive clusters.
Note that the dilution observed is also compatible with the one expected from the model M1 (cf. Table~\ref{table:Y}).




\begin{acknowledgements}
We acknowledge the Archeops collaboration for the use of
the proprietary Archeops data and related software as well as
for fruitful comments and careful reading of this manuscript.
We acknowledge very useful comments by J.A.Rubi\~no--Mart\1n.
C.H.M. acknowledges the financial support
from the European Community through the Human Potential
Programme under contract HPRN-CT-2002-00124 (CMBNET).
Some of the results in this paper have been derived using the HEALPix
package, \cite{healpix}.
We acknowledge the use of the Legacy Archive for Microwave
Background Data Analysis (LAMBDA, {\tt http://lambda.gsfc.nasa.gov}).
Support for LAMBDA is provided by the NASA Office of Space Science.
This publication makes use of data products from the Two Micron All Sky
Survey, which is a joint project of the University of Massachusetts and
the Infrared Processing and Analysis Center/California Institute of
Technology, funded by the National Aeronautics and Space Administration
and the National Science Foundation.

\end{acknowledgements}


\bibliography{aamnem99,paper}

\end{document}